\documentclass[12pt,a4paper]{article}
\usepackage{amsmath,amssymb,amsthm}
\usepackage{geometry}
\usepackage{booktabs}
\usepackage{array}
\usepackage{setspace}
\usepackage{natbib}
\usepackage{graphicx}
\usepackage{float}
\usepackage{needspace}
\usepackage{hyperref}
\hypersetup{hidelinks}

\newtheorem{proposition}{Proposition}

\newtheorem{corollary}{Corollary}

\title{Too fast to adjust\\
AI adoption speed and finite adjustment capacity}

\author{Eduardo Levy Yeyati\thanks{Universidad Torcuato Di Tella and Inter-American Development Bank. The views expressed in this paper are those of the author and do not necessarily reflect the views of the Inter-American Development Bank.}}

\begin{document}
\begingroup
\renewcommand{\thefootnote}{\fnsymbol{footnote}}
\begin{singlespace}
\begin{center}
\vspace*{-0.45in}
{\LARGE\bfseries Too fast to adjust\par}
\vspace{0.35em}
{\Large AI adoption speed and finite adjustment capacity\par}
\vspace{0.85em}
{\large Eduardo Levy Yeyati\footnotemark\par}
\vspace{0.35em}
{\large November 2026\par}
\end{center}
\footnotetext{Corresponding author: Eduardo Levy Yeyati, Universidad Torcuato Di Tella and Inter-American Development Bank. e-mail: ely@utdt.edu. The views expressed in this paper are those of the author and do not necessarily reflect the views of the Inter-American Development Bank.}

\vspace{0.55em}
\begin{abstract}
\noindent
Finite retraining capacity makes adoption speed matter even when eventual automation is fixed. Faster diffusion concentrates displacement, lengthens retraining queues, delays reemployment, and pushes some workers out of the labor force. I model exit as a stopping decision and use earnings-based measures of market output and welfare that differ in their treatment of nonwork. Shared retraining capacity also creates an externality: each additional displacement slows the reallocation of other workers. Under the reference capacity scenario, shortening the adoption transition by 70 percent raises the share of displaced workers persistently leaving the labor force from 8.0 to 26.0 percent. The acceleration requires a remaining labor productivity level gain at full adoption of 7.2 percent to break even in welfare and 12.0 percent in market output.
\end{abstract}

\end{singlespace}
\endgroup
\setcounter{footnote}{0}

\newpage

\section{Introduction}

Most analyses of automation ask how much work a technology will eventually perform. This paper holds the endpoint fixed and asks how fast the economy gets there. Speed matters when displaced workers have to pass through a retraining system of limited capacity: a worker who would retrain if the system were slack may leave the labor force when the queue is long.

Generative AI is a natural case. Its adoption in the workplace has been faster than that of personal computers, and its applications span a broad range of tasks \citep{bick2025}. Recent assessments emphasize the role of institutional readiness in technological transitions \citep{barhoumi2026,iaisafety2026}. The same logic applies to other fast-diffusing technologies that displace labor.

Retraining uses scarce inputs: instructors, training slots, credentials, counseling, mobility support, and time. Governments and firms can add capacity, but not instantly. A displacement wave that arrives twice as fast need not be processed twice as fast. As the adjustment queue grows relative to installed capacity, individual completion rates fall.

In the calibrated transitions, faster adoption lowers the value of remaining in adjustment through two channels. A longer queue delays retraining and reemployment. Successful reallocations also arrive in a more concentrated wave, temporarily lowering wages in the sector that absorbs reallocated workers. Workers can remain in adjustment or leave the labor force for an available social-protection option. I call the latter supported nonparticipation. For prime-age workers, the option can be interpreted as disability insurance; at older ages, retirement plays a similar role. The stopping decision depends on the combined private value of transfers, nonworking time, and access costs.

The paper's contribution is to show how faster adoption, by creating a temporary bottleneck in a system of finite adjustment capacity, raises the incidence of persistent labor-force exit among displaced workers, and to quantify the implications for alternative adoption speeds. Congestion increases transition costs through both forgone earnings during adjustment and the prospective productive stream lost when workers exit. It also creates an externality because each worker takes her completion rate as given while additional displacement affects other users of shared capacity. The welfare calculation credits the real value of nonworking time after exit.

The main quantitative exercise compares two adoption paths that take 14.6 and 4.4 years to move from 10 to 90 percent of eventual adoption, a reduction in transition duration of 70 percent. Under the reference capacity scenario, defined by the peak displacement flow under the slower path (2.625 percent of the workforce per year), the share of displaced workers who persistently exit the labor force rises from a calibrated benchmark of 8.0 percent to 26.0 percent. Under the earnings-based accounting criteria used here, the faster transition requires a remaining level gain in labor productivity at full adoption of 7.2 percent to break even in welfare and 12.0 percent in market output. Time in adjustment accounts for 3.8 percentage points of the market-output threshold, of which 3.4 reflect the net excess adjustment loss. This component is unaffected by the value assigned to nonwork after exit.

The relationship between capacity and the break-even gain is nonmonotonic. As capacity tightens, the welfare threshold rises from 3.4 to 9.0 percent before falling to 7.6 percent at the lowest capacity considered. When capacity is sufficiently scarce, many workers already leave under the slower transition, reducing the incremental cost of acceleration.

Holding destination wages and successful-reallocation values fixed, one additional displacement at the peak of the reference transition imposes on other users of adjustment capacity a loss equivalent to 1.29 years of average earnings, discounted to the beginning of the transition. Five years later, the corresponding loss is 0.46. The difference reflects both changing adjustment conditions and discounting. Under the fixed-price and common-gain assumptions of Corollary~\ref{cor:timing}, it creates a range of private gains for which earlier adoption is privately profitable but socially excessive.

The paper is closest to \citet{berajazorzi2025}, who show that automation can be too fast when displaced workers reallocate slowly and face borrowing constraints. \citet{lehrrestrepo2022} derive welfare gains from gradual technology and trade shocks in the presence of adjustment frictions, while \citet{adaoberajapandalai2024} study how skill specificity and cohort replacement shape technological transitions. \citet{guerreiro2022} and \citet{costinotwerning2023} examine the optimal regulation of automation more broadly. The average--marginal wedge created by shared congestion is familiar from queueing and matching models \citep{naor1969,hosios1990}. Here it arises in retraining capacity, interacts with endogenous labor-force exit, and is quantified for alternative adoption speeds. \citet{crepon2013} document a related externality in labor-market assistance operating through competition among job seekers.

Task-based models study how automation displaces labor and how new tasks restore labor demand in the long run \citep{acemoglu2018,acemoglu2022}. Recent work distinguishes automation, worker complementarity, and the creation of new human tasks, and examines how those choices shape private incentives \citep{acemogluautorjohnson2026}. Firm-level evidence from France shows that AI adoption can increase employment on average, with effects that depend on how firms use the technology \citep{aghion2025france}. A separate literature studies how quickly technologies diffuse \citep{comin2010,brynjolfsson2017}. Evidence from trade liberalization and import competition documents reallocation processes that last well beyond the initial shock \citep{autor2013china,dix2014,caliendo2019}.

Labor-force participation responds to labor-market conditions \citep{elsby2015}. The interpretation of supported nonparticipation draws on evidence linking adverse labor-demand shocks and job loss to disability entry and retirement \citep{autorduggan2003,bratsberg2013,foote2019}. Search models have long allowed unemployment spells to end in nonparticipation \citep{vandenberg1990}. Evidence from unemployment-insurance extensions shows that increasing the value of remaining attached reduces labor-force withdrawal, with little effect on job finding \citep{valletta2014,farber2015}. This evidence disciplines the stopping rule, while data on displaced workers provide the benchmark exit rate.

The valuation of transition losses draws on evidence of persistent earnings losses after displacement \citep{jacobson1993,stevens1997,couch2010,chanstevens2001}. The accounting excludes the job- and industry-specific capital lost at displacement and focuses on the productive capital transferable to another job \citep{neal1995,jacobson1993}. Training records provide a benchmark for adjustment timing, and recent evidence on AI-exposed workers suggests that formal retraining can facilitate reallocation \citep{hyman2025}. Following \citet{chodorowreichkarabarbounis2016}, the welfare calculation distinguishes transfers from the real value of nonworking time, drawing on the experimental estimates of \citet{maspallais2019}.

\section{Model}
\label{sec:model}

\subsection{Adoption and displacement}

The reference workforce is normalized to one. A fixed share $\bar\imath\in(0,1)$ of labor tasks (expressed in worker-equivalent units) is eventually automated. Let $a(t)\in[0,1]$ be the fraction of that envelope already adopted. Diffusion follows
\begin{equation}
 \dot a(t)=\kappa a(t)[1-a(t)],\qquad a(0)=a_0,
 \label{eq:adoption}
\end{equation}
so
\begin{equation}
 a(t;\kappa)=\frac{1}{1+[(1-a_0)/a_0]e^{-\kappa t}}.
\end{equation}
Adoption may be lumpy within individual firms or production units. The smooth path above describes the aggregate share that has switched technology. The displacement flow is
\begin{equation}
 \Phi(t;\kappa)=\bar\imath\kappa a(t;\kappa)[1-a(t;\kappa)].
 \label{eq:phi}
\end{equation}
For $a_0<1/2$, its peak is $\bar\imath\kappa/4$, while total remaining displacement is independent of speed:
\begin{equation}
 \int_0^\infty \Phi(t;\kappa)dt=\bar\imath(1-a_0).
 \label{eq:mass}
\end{equation}
Changing $\kappa$ changes only the timing and concentration of displacement, not its eventual amount.

\subsection{Finite adjustment throughput}

Let $U(t)$ be the stock of displaced workers still in adjustment, $R(t)$ cumulative successful retraining, and $D(t)$ cumulative exit from the labor force. Successful adjustments occur at rate
\begin{equation}
 M(U;Z)=Z\left(1-e^{-\rho_0 U/Z}\right),
 \label{eq:M}
\end{equation}
where $Z>0$ is installed adjustment capacity, measured in workers per unit of time, and $\rho_0>0$ is the completion hazard when the system is uncongested. The individual completion hazard is
\begin{equation}
 r(U,Z)=\begin{cases}M(U;Z)/U,&U>0,\\ \rho_0,&U=0.\end{cases}
 \label{eq:r}
\end{equation}
With a short queue, completions are approximately proportional to the number of workers in adjustment. With a long queue, total completions approach the ceiling $Z$.

\begin{proposition}[Finite-throughput congestion]\label{prop:throughput}
For $U,Z>0$, $M_U>0$, $M_Z>0$, $r_U<0$, and $r_Z>0$. Moreover,
\[
 \lim_{U/Z\to0}r(U,Z)=\rho_0,\qquad
 \lim_{U/Z\to\infty}\frac{M(U;Z)}{Z}=1.
\]
\end{proposition}

Write $r=\rho_0 f(x)$, where $x=\rho_0U/Z$ and $f(x)=(1-e^{-x})/x$. Since $f'(x)<0$ for $x>0$, a longer queue lowers each worker's completion hazard and more capacity raises it.

The flow equations are
\begin{align}
 \dot U(t)&=\Phi(t;\kappa)-M(U(t);Z)-\mu(t)U(t), \label{eq:dU4}\\
 \dot R(t)&=M(U(t);Z), \label{eq:dR4}\\
 \dot D(t)&=\mu(t)U(t). \label{eq:dD4}
\end{align}
The stocks count only displacement after the initial date: $U(0)=R(0)=D(0)=0$ for finite speeds, so the total remaining displacement is $\bar\imath(1-a_0)$. Adjustment associated with earlier adoption is outside the modeled transition. Every newly displaced worker eventually completes adjustment, counted by $M$, or exits.

\subsection{Destination-sector wages and the value of successful adjustment}

In the baseline, the sector that absorbs reallocated workers is technologically separate from the automating sector. Its inverse labor demand is
\begin{equation}
 w^N(t)=w_0^N\left[\frac{1-\bar\imath}{L^N(t)}\right]^{1/\sigma},
 \qquad L^N(t)=(1-\bar\imath)+R(t),
 \label{eq:w4}
\end{equation}
with $\sigma>0$ and $w_0^N=1$. The destination wage $w^N(t)$ is a dimensionless index relative to its initial level; earnings are measured in units of baseline mean annual earnings. Automation elsewhere therefore does not directly raise the marginal product of labor in this sector. The destination sector stands in for all activities that absorb reallocated workers, including non-automating sectors and units that have already adopted the new technology. The parameter $\sigma$ governs how easily those activities absorb additional labor. A faster inflow lowers the destination wage along this demand curve; if enough workers later exit, the wage ordering can reverse.

Let $J^N(t)$ be the discounted earnings value of completing adjustment at time $t$:
\begin{equation}
 J^N(t)=E_t\left[\int_t^{t+\mathcal T}e^{-\beta(s-t)}\widetilde w(s)\,ds\right],
 \label{eq:J4}
\end{equation}
where $\widetilde w$ is post-displacement earnings and $\mathcal T$ is a standardized prospective earnings horizon measured from the date of resolution. In the deterministic numerical model, expectations are perfect foresight and the value is constructed as
\begin{equation}
 J^N(t)=\sum_j s_j w_j^R\int_0^{\mathcal T_j}e^{-\beta v}w^N(t+v)\,dv,
 \qquad \mathcal T_j=-\frac{\log(1-\beta R_j^{PV})}{\beta}.
 \label{eq:Jimplementation}
\end{equation}
Here $w_j^R$ is the group-specific earnings scale relative to mean earnings, $s_j$ is its displaced-worker weight, and $R_j^{PV}$ is the discounted remaining-work factor in Appendix B. The earnings scale incorporates the group-specific retention factor; the numerical model does not impose an additional time-varying recovery profile. This is a representative-worker valuation with fixed age weights and rolling prospective horizons: the same $\mathcal T_j$ is used at each potential date of reemployment or exit, regardless of the time already spent in adjustment. The model does not follow individual aging or impose retirement dates at displacement. In particular, waiting does not shorten the prospective stream; for long spells this can overstate lost lifetime earnings relative to a fixed retirement horizon.

Let $\widehat V_t^p$ be the private value of staying in adjustment, normalized by the common earnings benchmark $\bar H$, and let $\widehat J_t^N=J_t^N/\bar H$. Opportunities to claim social protection arrive at Poisson rate $\chi$. When an opportunity arrives, the worker observes a private exit value $d$. This value can include an expected transfer, nonwork time, and application or access costs. The stopping decision depends only on their combined value, so the model places a distribution directly on $d$.

Conditional on remaining attached, the worker's Bellman equation is
\begin{equation}
 \beta \widehat V_t^p=\widehat u_U(t)+r(U(t),Z)[\widehat J_t^N-\widehat V_t^p]
 +\chi E[(d-\widehat V_t^p)_+]+\dot{\widehat V}_t^p,
 \label{eq:Vprivate}
\end{equation}
where the baseline sets $\widehat u_U=0$. Temporary unemployment-insurance payments enter $\widehat u_U$ in the calibration exercise below.

\subsection{Persistent exit as optimal stopping}

The worker exits whenever $d\geq\widehat V_t^p$. If $G$ is the distribution of private exit values, the exit hazard is
\begin{equation}
 \mu(t)=\chi[1-G(\widehat V_t^p)].
 \label{eq:mugeneral}
\end{equation}
For a logistic distribution of $d$ with location $\gamma/\lambda$ and scale $1/\lambda$,
\begin{align}
 \mu(\widehat V_t^p)&=\frac{\chi}{1+\exp[-(\gamma-\lambda\widehat V_t^p)]}, \label{eq:mulogit}\\
 \chi E[(d-\widehat V_t^p)_+]&=\frac{\chi}{\lambda}\log\left[1+\exp(\gamma-\lambda\widehat V_t^p)\right]. \label{eq:optionvalue}
\end{align}
With a logistic $G$, the stopping rule delivers the bounded logistic hazard in equation~\eqref{eq:mulogit}. The UI evidence supplies a target for its slope, matched to a stock-weighted average along the reference transition,
\begin{equation}
 s_V\equiv\frac{\int_0^{T_s} U(t)\left[-\partial\mu(\widehat V_t^p)/\partial\widehat V^p\right]dt}{\int_0^{T_s} U(t)\,dt},
 \label{eq:sV}
\end{equation}
where $T_s=100$ years in the numerical calibration and the paths are those of the reference transition. Conditional on $\chi$, $(\gamma,\lambda)$ are chosen to match this slope and the 7.97 percent benchmark exit share. The arrival rate $\chi$ is a normalization; Section~\ref{sec:robustness} shows that the results are insensitive to it.

The two channels, over intervals in which congestion and wage compression reduce the private value of successful adjustment, are:
\begin{equation*}
 \begin{aligned}
 &\kappa\uparrow\ \Rightarrow\ \Phi\text{ more concentrated}\ \Rightarrow\\[-1pt]
 &\qquad\begin{cases}
 U/Z\uparrow\ \Rightarrow\ r(U,Z)\downarrow &\text{(congestion)}\\
 \dot R\text{ concentrated}\ \Rightarrow\ w^N\downarrow\ \Rightarrow\ J^N\downarrow &\text{(destination absorption)}
 \end{cases}\\[3pt]
 &\qquad\Rightarrow\ \widehat V^p\downarrow\ \Rightarrow\ \mu\uparrow.
 \end{aligned}
\end{equation*}

The effect on cumulative exit depends on the joint paths of $\mu(t)$ and $U(t)$; the numerical comparisons below show the net effect for the calibrated transitions.

\section{Welfare and market output}
\label{sec:surplus}

Let $g_A$ be the remaining proportional increase in labor productivity between initial adoption $a_0$ and full adoption. For example, $g_A=0.07$ represents a 7 percent level gain over the initial state, not an annual growth rate. The labor share $\alpha_L$ converts that gain into the same wage-based output units as the transition costs. Define
\begin{equation}
 B(\kappa)=\frac{1}{\alpha_L}\int_0^\infty e^{-\beta t}
 \frac{a(t;\kappa)-a_0}{1-a_0}\,dt.
 \label{eq:B}
\end{equation}
The term $g_AB(\kappa)$ is the discounted gain from reaching the same final productivity level earlier. The model defines $g_A$ net of technology services, complementary capital, implementation, and reorganization costs. The external magnitudes below are treated as illustrative values of this net gain, rather than direct estimates of it.

If a worker exits at time $t$, the accounting assigns a productive loss equal to her standardized prospective earnings value at that date. Using market earnings as a proxy for marginal product, define
\begin{equation}
 C_D(\kappa;Z)=\int_0^\infty e^{-\beta t}J^N(t)\mu(t)U(t)dt.
 \label{eq:CD}
\end{equation}
$C_D$ counts only productive capacity that can be carried into a new job. Job- and industry-specific capital lost at displacement is excluded by construction; although its undiscounted loss may be invariant to speed, its discounted value need not be \citep{jacobson1993,neal1995}. Because $J^N(t)$ is the value of successful reallocation at date $t$, an exit is valued using the productive opportunities available when it occurs.

Finite capacity also makes successful reallocation take longer. Let $\bar w_R=\sum_j s_jw_j^R$ denote the average post-displacement earnings scale used in the CPS calibration and define
\begin{equation}
 C_U(\kappa;Z)=\int_0^\infty e^{-\beta t}\bar w_R w^N(t)U(t)dt.
 \label{eq:CU}
\end{equation}
The two terms cover different parts of a worker's path. $C_U$ accrues only while she remains in adjustment. If she exits, $C_D$ starts at the exit date and values the market stream she would otherwise have produced. The two losses do not overlap in calendar time, but their sum is not a fixed-lifetime earnings loss: the standardized prospective horizon starts at exit, as specified above.

Workers would spend some time retraining even with unlimited aggregate capacity. To measure the net excess adjustment-time loss associated with congestion, let $C_U^0(\kappa)$ be the adjustment loss in a counterfactual with proportional throughput, $M^0(U)=\rho_0U$, after re-solving the stopping and destination equations. The excess adjustment-time loss is
\begin{equation}
 C_Q(\kappa;Z)=C_U(\kappa;Z)-C_U^0(\kappa).
 \label{eq:CQ}
\end{equation}
This counterfactual difference includes changes in time spent in adjustment, exit behavior, and the wages used to value that time. Both ordinary and excess adjustment losses enter the aggregate accounting below. Define the discounted market-output criterion, relative to the initial productivity baseline, as
\begin{equation}
 \mathcal Y(\kappa;Z,g_A)=g_AB(\kappa)-C_D(\kappa;Z)-C_U(\kappa;Z).
 \label{eq:Y}
\end{equation}

This is an earnings-based partial-equilibrium accounting criterion, not the exact difference between aggregate production paths. In particular, forgone production is valued at the simulated earnings index rather than by integrating the destination sector's marginal-product schedule over employment differences. The technological dividend is imposed separately. For welfare, public transfers are consolidated, but time outside employment still has real value. Let $\xi\in[0,1]$ be that nonmarket value as a fraction of the productive stream forgone after exit. Welfare is
\begin{equation}
 \mathcal W(\kappa;Z,g_A,\xi)=g_AB(\kappa)-(1-\xi)C_D(\kappa;Z)-C_U(\kappa;Z).
 \label{eq:W}
\end{equation}
This is a risk-neutral social accounting criterion: displaced workers receive no extra distributional weight. It is not derived from separately identified private utility components. The parameter $\xi$ affects only the social value assigned to an observed exit; it does not change the stopping rule. Section~\ref{sec:nonwork} takes its baseline value from external evidence. Following \citet{chodorowreichkarabarbounis2016}, transfers are treated as transfers while nonworking time counts as real value. The benchmark does not add an excess burden from financing social protection.

Nonmarket value is credited only after a worker leaves the labor force. Time in $U$ is spent searching, retraining, obtaining credentials, or otherwise trying to reenter employment, so the baseline does not assign it the same value. If time in the queue also has nonmarket value, the model overstates the welfare cost of waiting.

Appendix~\ref{app:frontier} characterizes the instantaneous- and slow-adoption limits. The quantitative analysis below focuses on comparisons between finite adoption speeds.

\subsection{Congestion on common adjustment capacity}
\label{sec:externality}

Because total completions are capped by capacity, when $U>0$ one more worker in the queue adds less to aggregate completions than the average worker already there:
\begin{equation}
 r(U,Z)-M_U(U,Z)=-U\,r_U(U,Z)=\rho_0\!\left[f(x)-e^{-x}\right]>0,
 \qquad x=\rho_0 U/Z,
 \label{eq:crowdout}
\end{equation}
with $f(x)=(1-e^{-x})/x$. This gap is the contemporaneous congestion imposed on other workers sharing the same capacity. Congestion keeps them in adjustment longer and, by lowering the value of staying attached, can also induce additional exit.

To measure the marginal cost of displacement while isolating the common-capacity channel, define
\begin{equation}
 \Lambda^W_\Phi(t)\equiv \frac{\delta C^W}{\delta \Phi(t)},
 \qquad C^W=(1-\xi)C_D+C_U,
 \label{eq:gateauxW}
\end{equation}
holding the destination wage $w^N$ and successful-reallocation value $J^N$ paths fixed, but allowing the private stopping value and exit hazard to respond. Let $L^{\rm own,W}(t)$ be the social loss attributable to the marginal worker's own transition, taking the aggregate paths $r$, $\mu$, $w^N$, and $J^N$ as given. With $S(s\mid t)=\exp[-\int_t^s(r(v)+\mu(v))\,dv]$ denoting survival in adjustment,
\begin{equation}
 L^{\rm own,W}(t)=\int_t^\infty e^{-\beta s} S(s\mid t)
 \left[\bar w_R w^N(s)+(1-\xi)J^N(s)\mu(s)\right]ds.
 \label{eq:ownloss}
\end{equation}
This is a social loss attributed to one worker, not necessarily her private utility loss. The congestion externality on other users, conditional on the fixed paths specified above, is
\begin{equation}
 \mathcal E^W(t)=\Lambda^W_\Phi(t)-L^{\rm own,W}(t).
 \label{eq:externalW}
\end{equation}
Under perfect foresight, this derivative includes effects on workers already attached, subsequent entrants, and workers who change their stopping behavior in anticipation of future congestion. It is not a cohort-specific measure or the full equilibrium welfare derivative.

\Needspace{12\baselineskip}
\begin{proposition}[Social protection and congested adjustment]\label{prop:socialprotection}
Suppose successful reallocation has positive productive value and the private exit value is summarized by $d$. Then: (i) the stopping problem in \eqref{eq:Vprivate}--\eqref{eq:mulogit} generates the aggregate exit hazard; (ii) under finite throughput and $U>0$, $0<M_U<r$, so an attached worker takes a completion rate as given that exceeds her marginal contribution to aggregate completions; and (iii) a marginal displacement decomposes into the worker's own transition loss and a congestion externality on other users, $\Lambda^W_\Phi=L^{\rm own,W}+\mathcal E^W$. The externality vanishes under proportional throughput $M(U)=\rho_0U$. Holding exit behavior fixed, the component of $\mathcal E^W$ due to additional waiting is positive whenever added load lowers $r$; the sign of the total externality, which also includes induced exit, is not determined analytically.
\end{proposition}

An exit need not be inefficient. The worker chooses whether to remain in adjustment, while the externality arises because retraining capacity is shared. A lower destination wage affects that private choice by reducing the return to successful reallocation. Under competitive pricing, however, wage compression is not counted as a separate social loss.

\begin{corollary}[Finite-date adoption with unpriced capacity]\label{cor:timing}
Let an adopter choose whether to cause a marginal displacement at date $t$ or at a later date $T>t$. Hold the destination wage and successful-reallocation value paths fixed, and suppose that private and social gains from earlier adoption coincide before worker-side transition costs are included. Let $\Delta\Pi_{tT}$ denote the discounted private gain from adopting at $t$ rather than $T$, before worker-side transition losses, and let $\Delta L_{tT}=L^{\rm own,W}(t)-L^{\rm own,W}(T)$, $\Delta\mathcal E_{tT}=\mathcal E^W(t)-\mathcal E^W(T)$, and $\Delta\Lambda_{tT}=\Lambda^W_\Phi(t)-\Lambda^W_\Phi(T)$. If the adopter bears the displaced worker's own transition loss, private early adoption requires $\Delta\Pi_{tT}\geq\Delta L_{tT}$ whereas social early adoption requires $\Delta\Pi_{tT}\geq\Delta\Lambda_{tT}$; whenever $\Delta\mathcal E_{tT}>0$, there is therefore a nonempty range of private gains for which the adopter chooses $t$ and the planner chooses $T$. If the adopter bears none of the worker-side transition loss, private early adoption requires $\Delta\Pi_{tT}\geq0$ while social early adoption requires $\Delta\Pi_{tT}\geq\Delta\Lambda_{tT}$; whenever $\Delta\Lambda_{tT}>0$, the interval $0\leq\Delta\Pi_{tT}<\Delta\Lambda_{tT}$ yields the same divergence.
\end{corollary}

Section~\ref{sec:externality_results} quantifies these wedges in the calibrated transition.

\section{Quantitative inputs}
\label{sec:calibration}

Table~\ref{tab:identification} summarizes the empirical source for each quantitative input.

\begin{table}[H]
\centering
\caption{Quantitative inputs}\label{tab:identification}
\small
\begin{tabular}{>{\raggedright\arraybackslash}p{.14\textwidth}>{\raggedright\arraybackslash}p{.31\textwidth}>{\raggedright\arraybackslash}p{.46\textwidth}}
\toprule
Object & Role & Empirical basis\\
\midrule
$s_V$ & Local response of persistent exit to the private continuation value & Quasi-experimental UI-duration evidence, mapped into $\widehat V^p$\\
$(\gamma,\lambda)$ & Location and scale of total private exit values & CPS cross-sectional exit proxy plus $s_V$, conditional on $\chi$\\
$\chi$ & Opportunity-rate normalization & Baseline $0.293$ per year; wide sensitivity reported below\\
$\xi$ & Real nonmarket value after exit & Baseline 0.58 from \citet{maspallais2019}; 0.30 lower and 0.72 upper sensitivities\\
$\bar\imath$ & Automation-envelope scenario & Baseline 0.35; sensitivity from 0.25 to 0.45\\
$\rho_0$ & Uncongested completion hazard & WIOA training completion followed by Q2 employment\\
$\bar H$ & Productive earnings at risk & CPS post-displacement earnings and remaining work horizons\\
$Z$ & Installed aggregate successful-adjustment throughput & Capacity scenarios in workforce-equivalent units per year\\
$g_A$ & Remaining level gain in labor productivity from initial to full adoption & External productivity scenarios\\
\bottomrule
\end{tabular}
\begin{minipage}{.94\textwidth}\footnotesize
Notes: $\bar\imath$, $Z$, and $g_A$ are scenario parameters. The value of $\xi$ comes from external evidence and affects welfare accounting only; it does not enter the stopping calibration.
\end{minipage}
\end{table}

\subsection{Displacement, earnings, and productive value}

The baseline sets the eventual automation envelope at $\bar\imath=0.35$ and initial adoption at $a_0=0.10$. The discount rate is $\beta=0.04$, the reference destination labor-demand elasticity is $\sigma=1.5$, and the labor share used to value the technology gain is $\alpha_L=0.60$. Larger values of $\sigma$ are considered in the robustness checks.

The January 2024 CPS Displaced Worker Supplement provides age shares, post-displacement earnings, and a benchmark measure of exit \citep{census2024}. Among the 1,425 displaced workers aged 20--67, the age shares weighted by the CPS supplement weight are $(0.3777,0.4138,0.2085)$ for ages 20--34, 35--54, and 55--67.

To construct the exit measure, start with the official displaced-worker sample (\texttt{PRDISPWK}=1) and use the supplement weight \texttt{PWSUPWGT}. A respondent is counted as a behavioral exit if she is currently out of the labor force (\texttt{PEMLR}=5--7) and reports either unemployment-insurance receipt (\texttt{PESD20}) or paid work since the lost job ended (\texttt{PESD24}). Current employment (\texttt{PEMLR}=1--2) and these exits are treated as resolved episodes; current unemployment (\texttt{PEMLR}=3--4) remains unresolved and is left out of the denominator. The resulting exit rates are 3.95, 6.41, and 18.34 percent across the three age groups, for an age-standardized average of 7.97 percent using the displaced-worker shares given above. This is the calibration target, not a claim about the pooled rate among resolved episodes. Because the supplement is a cross-section, this is a benchmark exit proxy, not a longitudinal measure of persistent nonparticipation. It can include withdrawal after reemployment, whereas the modeled exit occurs before successful reallocation.

Combining post-displacement earnings with discounted work factors yields
\begin{equation}
 \bar H=13.44
\end{equation}
average annual-wage-years. This scale normalizes $\widehat V^p$. The empirical exit rate rises with age. The baseline does not model age-specific stopping or changes in the age composition of the queue; it uses the displaced-worker weights and work factors to construct a common valuation scale.

\subsection{Uncongested completion and aggregate capacity}

WIOA records provide a benchmark for how quickly a formal training episode can be completed when aggregate capacity is not binding. Among dislocated workers, annual hazards for training completion followed by employment in the second quarter after exit are approximately 1.37, 1.27, and 1.07 across the three age groups. Weighting by the CPS displaced-worker shares gives
\begin{equation}
 \rho_0=1.26\ \text{per year},
 \label{eq:rhoemp}
\end{equation}
which is the baseline uncongested completion hazard. WIOA participants are selected, and formal training is only one route back to work, so $\rho_0$ is used to anchor timing rather than to measure economy-wide adjustment capacity \citep{dolwioa2025,hyman2025}.

Aggregate capacity $Z$ is not estimated from WIOA. Results are instead reported for a range of capacity levels, measured in worker-equivalents per year. For scale, 37,119 dislocated-worker participants received WIOA training in Program Year 2024, although WIOA is only one adjustment channel \citep{dolwioaperf2025}. At the reference adoption speed $\kappa=0.30$, peak displacement is
\begin{equation}
 \Phi_{\max}(0.30)=\frac{0.35\times0.30}{4}=0.02625.
\end{equation}
I therefore summarize capacity by the reference load $\ell=\Phi_{\max}(0.30)/Z$: the peak displacement flow under $\kappa=0.30$ relative to annual capacity.

\begin{table}[H]
\centering
\caption{Interpreting installed aggregate adjustment capacity}\label{tab:capacity}
\begin{tabular}{ccc}
\toprule
Reference load $\ell$ & $Z$ & Maximum annual throughput\\
\midrule
0.5 & 0.0525 & 5.25\% of workforce\\
1.0 & 0.0263 & 2.63\% of workforce\\
2.0 & 0.0131 & 1.31\% of workforce\\
4.0 & 0.0066 & 0.66\% of workforce\\
6.0 & 0.0044 & 0.44\% of workforce\\
\bottomrule
\end{tabular}
\begin{minipage}{.92\textwidth}\footnotesize
Notes: $Z$ is all-channel successful adjustment throughput. The table gives the economic scale of the capacity scenarios.
\end{minipage}
\end{table}

\subsection{Local participation response from unemployment insurance}

The UI evidence pins down how exit responds when the value of remaining attached changes. For UI-eligible workers unemployed at least 26 weeks in the 2007--11 matched-CPS sample, the monthly total unemployment-exit rate is $p_U=0.171$: $0.078$ to employment and $p_N=0.093$ to nonparticipation \citep[Table~2]{valletta2014}. For the value of an additional week of UI, any exit ends the current unemployment spell, so the total rate is the relevant discounting hazard. For the exit decision, however, the relevant outcome is withdrawal from the labor force. Ten additional weeks of potential benefits change that exit probability by $-0.00514$; the minus sign is established in the published erratum \citep[Table~5, Panel~B]{valletta2014,valletta2015erratum}. \citet{farber2015} similarly find little effect on job finding and a reduction in labor-force exit when extended benefits are available.

Convert the total unemployment-exit probability into an annual hazard and the two labor-force-exit probabilities into probability-equivalent annualized hazard proxies (not exact cause-specific hazards with competing risks),
\begin{equation}
 \begin{aligned}
 h_U&=-12\log(1-p_U),\qquad q_0=-12\log(1-p_N),\\
 q_1&=-12\log[1-(p_N-0.00514)].
 \end{aligned}
\end{equation}
For an assumed replacement rate $b=0.52$ and a ten-week extension $\tau_{UI}=10/52$, valued as if the additional payment stream began immediately rather than after remaining entitlement expires, the increase in the continuation value is approximated by
\begin{equation}
 \Delta V_{att}^{UI}=b\frac{1-e^{-(\beta+h_U)\tau_{UI}}}{\beta+h_U}.
 \label{eq:UIvalue}
\end{equation}
The calibration assumes that one quarter of the UI-induced change in labor-force withdrawal maps into the persistent exit state, so $p_D=0.25$. This is a bridge between the UI experiment and the model, not a claim that the CPS cross-section measures persistence. It implies the local slope
\begin{equation}
 s_V^{UI}=p_D\frac{q_0-q_1}{\Delta V_{att}^{UI}/\bar H}\simeq2.82.
 \label{eq:UIslope}
\end{equation}
\citet{krueger2014} provide supporting evidence that withdrawal among the long-term unemployed is persistent. I set the opportunity rate to $\chi=p_Dq_0\simeq0.293$ per year and choose $(\gamma,\lambda)$ so that the stopping rule matches the $U$-weighted average slope in equation~\eqref{eq:sV} of 2.82 and the CPS exit target $D(T_s)/[\bar\imath(1-a_0)]=0.07966$ at $\kappa=0.30$ and reference load $\ell=1$. This gives $(\gamma,\lambda)=(43.78,59.11)$ in normalized-value units. Appendix~\ref{app:numerics} repeats the calibration with opportunity rates two and four times as large.

The CPS does not reveal how congested the adjustment system was when the observed exits occurred. I therefore repeat the behavioral calibration at reference loads 0.5, 1, and 2 and evaluate all three calibrations on the same absolute $Z$ grid. This exercise shows how much the finite-speed results depend on the unknown congestion state behind the CPS moment.

\subsection{Value of nonwork}\label{sec:nonwork}

The exit calibration identifies when workers leave, but not how much real value they retain outside employment. \citet{maspallais2019} use randomized wage--hours offers to estimate the value of nonwork and report $\widetilde z=0.58$ (standard error $0.04$) relative to predicted pretax market earnings for a full-time job. Because $J^N$ is measured in market-earnings units and the destination sector is competitive, the baseline sets $\xi=0.58$. Changing $\xi$ changes welfare accounting only; it leaves the transition paths and the stopping decision unchanged. Appendix~\ref{app:numerics} also reports $\xi=0.30$ to allow for limited external validity and $\xi=0.72$ as an upper benchmark based on the broader Mas--Pallais opportunity-cost measure. Following \citet{chodorowreichkarabarbounis2016}, cash transfers affect private incentives but are consolidated in welfare, while nonworking time is treated as real value.
\subsection{Technological-gain scenarios}

I use three external estimates to gauge the size of the technology gain: $g_A=0.014$ from reported hours, and OECD lower and upper cases of $0.0407$ and $0.0937$ \citep{bick2025,filippucci2024}. The OECD numbers come from cumulating ten-year productivity-growth paths, approximately $1.004^{10}-1$ and $1.009^{10}-1$. The reported-hours and compounded OECD magnitudes are used as illustrative scenarios for the net remaining gain from the initial state to full adoption, not as estimates of that specific object. The external estimates do not separately identify how much of that gain had already been realized at $a_0$. Its timing is determined by $a(t;\kappa)$, so the OECD diffusion path is not counted twice. These studies do not jointly identify the productivity gain and the eventual automation envelope.

\section{Results}
\label{sec:results}

\subsection{Finite throughput creates the speed cost}

\begin{figure}[H]
\centering
\includegraphics[width=.70\textwidth]{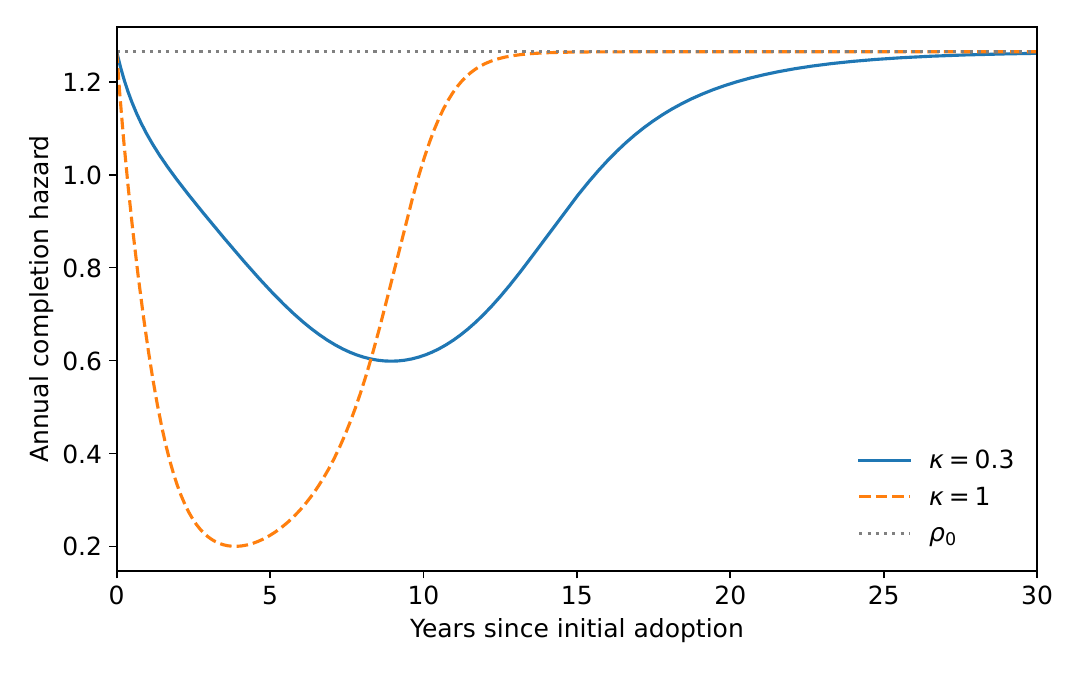}
\caption{Completion-hazard paths at reference load $\ell=1$. The horizontal line is the uncongested hazard $\rho_0$.}\label{fig:completion}
\end{figure}

Figure~\ref{fig:completion} shows what congestion does to an individual worker's completion rate at reference load $\ell=1$. Under the slower path, $\kappa=0.30$, the annual hazard already falls well below the uncongested rate of 1.26. Under $\kappa=1$, it falls sooner and farther. Faster adoption therefore lengthens the queue and slows the progress of each worker in it.

\begin{figure}[H]
\centering
\includegraphics[width=.68\textwidth]{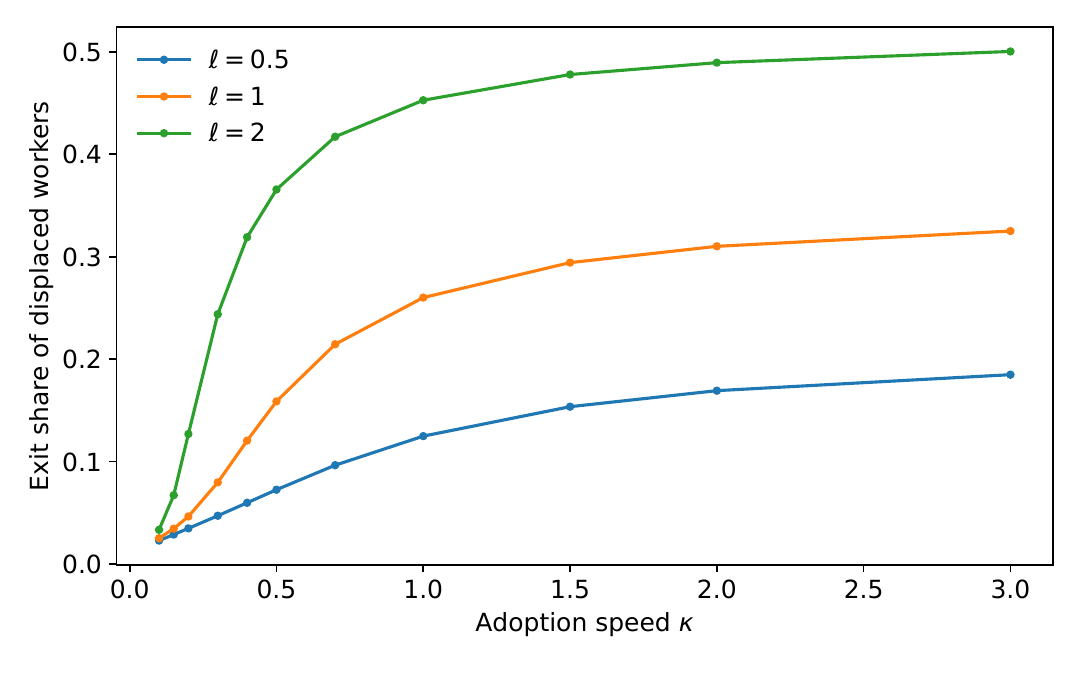}
\caption{Persistent exit as a share of displaced workers across adoption speeds and capacity levels under the stopping rule.}\label{fig:exit}
\end{figure}

With $\sigma=1.5$, accelerating from $\kappa=0.30$ to $1$ raises labor-force exit from 8.0 to 26.0 percent of displaced workers at load 1. At load 2, it rises from 24.4 to 45.3 percent. Figure~\ref{fig:exit} traces exit across speeds and capacity levels. The increase in exit from a given acceleration is not monotone in capacity. When capacity is scarce enough that many workers already leave under the slower path, faster adoption has less room to raise exit.

\subsection{Welfare and market-output thresholds}

The main welfare question is whether the productivity benefit from faster diffusion is large enough to offset the additional transition cost. For two finite speeds $\kappa_2>\kappa_1$, define the break-even gains
\begin{align}
 g^Y(\kappa_1,\kappa_2;Z)&=\frac{\Delta C_D+\Delta C_U}{\Delta B}, \label{eq:pairthresholdY}\\
 g^W(\kappa_1,\kappa_2;Z;\xi)&=\frac{(1-\xi)\Delta C_D+\Delta C_U}{\Delta B}. \label{eq:pairthresholdW}
\end{align}
Under the accounting criteria in Section~\ref{sec:surplus}, if the remaining productivity gain exceeds $g^Y$, the faster path raises the discounted output measure; if it exceeds $g^W$, it raises the welfare measure. Table~\ref{tab:pairwise} reports the main comparison, from $0.30$ to $1$, and a second acceleration from $1$ to $2$. The exit and adjustment-time columns show how the market-output threshold is built before the welfare value of nonwork is credited.

\begin{table}[H]
\centering
\caption{Pairwise welfare and market-output thresholds across installed capacity}\label{tab:pairwise}
\small
\resizebox{\textwidth}{!}{%
\begin{tabular}{cccccccccc}
\toprule
$Z$ (\%/yr) & Ref. load & \multicolumn{4}{c}{$0.30\to1$} & \multicolumn{4}{c}{$1\to2$}\\
\cmidrule(lr){3-6}\cmidrule(lr){7-10}
 & & $g^W(.58)$ & $g^Y$ & Exit & Adjustment & $g^W(.58)$ & $g^Y$ & Exit & Adjustment\\
\midrule
5.25 & 0.5 & 0.034 & 0.055 & 0.035 & 0.019 & 0.089 & 0.138 & 0.084 & 0.054\\
2.63 & 1.0 & 0.072 & 0.120 & 0.081 & 0.038 & 0.110 & 0.175 & 0.111 & 0.064\\
1.31 & 2.0 & 0.090 & 0.155 & 0.114 & 0.042 & 0.100 & 0.163 & 0.109 & 0.054\\
0.66 & 4.0 & 0.081 & 0.145 & 0.109 & 0.035 & 0.085 & 0.144 & 0.102 & 0.042\\
0.44 & 6.0 & 0.076 & 0.136 & 0.104 & 0.033 & 0.078 & 0.135 & 0.099 & 0.036\\
\bottomrule
\end{tabular}%
}
\begin{minipage}{.96\textwidth}\footnotesize
Notes: The exit and adjustment-time columns are the market-output components before the welfare weight is applied. The adjustment column includes ordinary training time. Totals can differ slightly due to rounding. Welfare uses $\xi=0.58$. The load is peak displacement at $\kappa=0.30$ divided by installed capacity $Z$.
\end{minipage}
\end{table}

\begin{figure}[H]
\centering
\includegraphics[width=.78\textwidth]{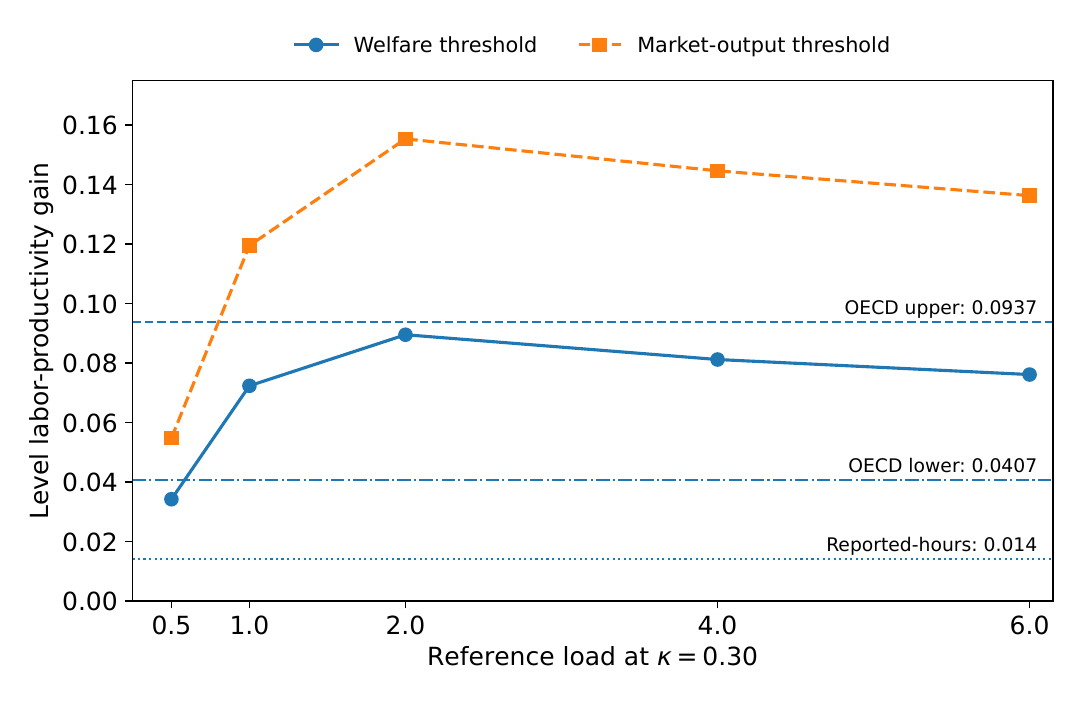}
\caption{Welfare and market-output thresholds for accelerating from $\kappa=0.30$ to $1$. Horizontal lines show the reported-hours, OECD lower, and OECD upper technological-gain scenarios.}\label{fig:thresholds}
\end{figure}

The relationship with capacity is hump-shaped. For the $0.30\to1$ acceleration, the welfare break-even gain rises from 3.4 percent at load 0.5 to 9.0 percent at load 2, then falls to 7.6 percent at load 6. The corresponding market-output threshold rises from 5.5 to 15.5 percent before falling to 13.6 percent. At the scarcest capacity levels, exit is already high under the slower path, so faster adoption adds less exit and less additional adjustment time. The $1\to2$ comparison flattens sooner for the same reason.

At reference capacity, compressing the transition by about 70 percent requires a remaining productivity level gain at full adoption of 7.2 percent to break even in welfare and 12.0 percent to break even in market output. Of the 12.0 percentage-point output threshold, 8.1 points come from additional exit and 3.8 from additional adjustment-time losses. The latter consists of 0.4 points of ordinary-training timing and 3.4 points of net excess adjustment loss. Crediting nonwork at $\xi=0.58$ reduces the exit contribution to welfare to about 3.4 points, but leaves the adjustment-time cost unchanged. For the further acceleration from $1$ to $2$, the break-even gains are 11.0 percent for welfare and 17.5 percent for market output. Adjustment time therefore remains an important part of the cost even under a generous valuation of nonwork.

Figure~\ref{fig:thresholds} compares these break-even gains with the three external productivity scenarios. Under the baseline welfare valuation, the OECD upper case, 9.37 percent, is large enough to justify the $0.30\to1$ acceleration at every capacity shown. The OECD lower case, 4.07 percent, clears the welfare threshold only when capacity is relatively slack. Market output sets a higher bar: at reference capacity even the OECD upper case falls short of the 12.0 percent threshold.

Appendix~\ref{app:frontier} reports the limiting comparison with instantaneous adoption.

\subsection{Congestion externality}\label{sec:externality_results}

Table~\ref{tab:external} measures the congestion externality at the displacement peak under $\kappa=0.30$. The calculation holds the destination wage and value paths fixed so that only the shared-capacity channel moves. Welfare uses the baseline $\xi=0.58$.

\begin{table}[H]
\centering
\caption{Congestion crowd-out and the marginal welfare cost of displacement}
\label{tab:external}
\begin{tabular}{lcccc}
\toprule
$Z$ (\%/yr) & $|\varepsilon_{rU}|$ & $\Lambda^W_\Phi$ & $L^{\rm own,W}$ & $\mathcal E^W$\\
\midrule
5.25 & 0.27 & 1.07 & 0.76 & 0.31\\
2.63 & 0.60 & 2.51 & 1.22 & 1.29\\
1.31 & 0.96 & 4.52 & 2.67 & 1.85\\
\bottomrule
\end{tabular}
\begin{minipage}{.94\textwidth}\footnotesize
Notes: Reference stopping calibration at $\kappa=0.30$, evaluated at the displacement peak. $|\varepsilon_{rU}|=(r-M_U)/r$ is the exact crowd-out elasticity of the throughput technology. $\Lambda^W_\Phi$ is the Gateaux derivative of $(1-\xi)C_D+C_U$ with respect to the displacement flow, in discounted average-worker annual-wage-year units per unit of workforce displaced, holding the destination wage/value path fixed. $L^{\rm own,W}$ is the marginal worker's own expected transition loss under the same criterion, and $\mathcal E^W=\Lambda^W_\Phi-L^{\rm own,W}$. Welfare uses $\xi=0.58$. Components may differ from totals by rounding.
\end{minipage}
\end{table}

At reference capacity, an additional worker displaced at the peak generates a total marginal welfare loss equal to 2.51 years of average earnings in present value. Of that loss, 1.22 is attributable to the displaced worker's own transition and 1.29 to effects on other users of adjustment capacity. Doubling capacity reduces that external loss to 0.31, while halving capacity raises it to 1.85.

Corollary~\ref{cor:timing} compares displacement at the peak, $t=7.3$, with displacement five years later, $T=12.3$. Over that interval, the externality falls from 1.29 to 0.46 years of average earnings, a difference of 0.83. The total marginal worker-side cost falls from 2.51 to 1.22, a difference of 1.29. \citet{crepon2013} find that labor-market assistance can also displace untreated job seekers. Their mechanism is job competition, not processing capacity, but the evidence shows that adjustment programs can generate spillovers across workers that matter economically.

\subsection{Robustness}\label{sec:robustness}

Table~\ref{tab:robustsummary} collects the main robustness checks for the $0.30\to1$ welfare comparison at reference capacity. Two assumptions matter most: the value assigned to nonwork and the eventual amount of automation. Changing the throughput function, the calibration load, destination-sector absorption, the persistence bridge, or the opportunity-rate normalization leaves the break-even gain close to 7.2 percent. If adjustment capacity grows by 5 percent per year, the threshold falls to 6.5 percent. Appendix~\ref{app:numerics} reports the full tables.

\begin{table}[H]
\centering
\caption{Welfare-threshold robustness at reference capacity}\label{tab:robustsummary}
\small
\begin{tabular}{lc}
\toprule
Alternative & $g^W(0.30,1)$\\
\midrule
Baseline & 0.072\\
Calibration-load band & 0.071--0.079\\
Michaelis--Menten throughput & 0.074\\
Capacity grows 5\% annually (recalibrated) & 0.065\\
$\xi=0.30$ / $0.72$ & 0.095 / 0.061\\
$\sigma=3$ / $6$ & 0.077 / 0.080\\
$p_D=0.15$ / $0.35$ & 0.074 / 0.071\\
$\chi\times4$ & 0.072\\
Automation envelope $\bar\imath=0.25$ / $0.45$ & 0.038 / 0.099\\
\bottomrule
\end{tabular}
\begin{minipage}{.90\textwidth}\footnotesize
Notes: All entries refer to the $0.30\to1$ welfare threshold at $Z=2.625$ percent of the workforce per year unless stated otherwise. The calibration-load band rematches the benchmark behavioral moments at reference loads $0.5$, $1$, and $2$. The capacity-growth entry uses the recalibrated stopping rule.
\end{minipage}
\end{table}

At reference capacity, changing the load at which the behavioral rule is calibrated moves the welfare threshold from 7.1 to 7.9 percent; the corresponding market-output range is wider, from 8.4 to 12.1 percent. The difference reflects the fact that only 42 percent of the exit loss enters welfare when $\xi=0.58$. Replacing the exponential throughput function with Michaelis--Menten gives a 7.4 percent welfare threshold. Raising $\sigma$ to 3 or 6 gives 7.7 and 8.0 percent.

The welfare valuation and the automation envelope move the result more. The threshold is 9.5 percent at $\xi=0.30$ and 6.1 percent at the upper benchmark $\xi=0.72$. Holding absolute capacity fixed, reducing the eventual automation share from 0.35 to 0.25 lowers the threshold to 3.8 percent; raising it to 0.45 increases the threshold to 9.9 percent. Once the behavioral moments are rematched, changing the opportunity-rate normalization or the persistence bridge has little effect. Appendix~\ref{app:numerics} contains the secondary $1\to2$ comparison and the numerical diagnostics.

\section{Conclusions}

The same amount of automation can be much harder to absorb when it arrives quickly. With limited retraining capacity, a faster transition creates a longer queue, slows the completion of training, and sends reallocated workers into the destination sector in a more concentrated wave. Staying in adjustment becomes less attractive, and more displaced workers leave the labor force.

The earnings-based accounting separates two costs of that acceleration. Some workers exit before they are reallocated, forgoing a prospective earnings stream. Time spent in adjustment by all attached workers also carries a cost, whether they eventually return to employment or exit. At benchmark capacity, shortening the adoption transition by about 70 percent requires a remaining productivity level gain at full adoption of 7.2 percent to break even in welfare and 12.0 percent in market output. Valuing nonwork more or less generously moves the welfare threshold between 6.1 and 9.5 percent. Net excess adjustment loss accounts for 3.4 percentage points of the market-output threshold and is unaffected by that valuation.

Shared retraining capacity also creates an externality. A worker displaced at the peak imposes on other users of shared adjustment capacity a present-value loss equal to 1.29 years of average earnings at benchmark capacity, holding the destination wage and reallocation-value paths fixed. Five years later that externality is 0.46. Under the assumptions of Corollary~\ref{cor:timing}, an adopter that does not pay for this difference can therefore prefer displacement during the peak even when waiting would raise social welfare.

The quantitative results also depend on the representative-worker earnings valuation rather than explicit aging and retirement. The main empirical limitation is that adjustment capacity is not measured. I treat aggregate capacity $Z$ as a scenario and impose, rather than estimate, the shape of the throughput technology. The next step is to measure how completion rates respond to load and how quickly retraining capacity can expand. The productivity gain needed to justify faster adoption depends on how crowded the adjustment system already is. A transition that pays off in a prepared labor market can be too fast in a congested one.

\section*{Data Availability Statement}

The analysis uses public CPS and WIOA data. The replication package includes the CPS extraction and calibration code, the archived WIOA spell-level extract and the original WIOA estimator, and the numerical model and diagnostics. CPS moments and the WIOA hazards from archived spells have been reproduced; extraction of the archived WIOA spells from the original public-use file has not been independently rerun in the current environment. The public-use source is identified in the replication instructions.

\bibliographystyle{plainnat}
\bibliography{refs}

\newpage
\section*{Appendices}
\appendix
\noindent\textit{Appendices A--E are intended for online publication.}\par
\vspace{0.5em}
\numberwithin{equation}{section}

\section{Proofs and analytical details}

\subsection{Proof of Proposition~\ref{prop:throughput}}
Let $x=\rho_0U/Z$. Then
\[
 M=Z(1-e^{-x}),\qquad r=\rho_0\frac{1-e^{-x}}{x}.
\]
Direct differentiation gives $M_U=\rho_0e^{-x}>0$ and $M_Z=1-(1+x)e^{-x}>0$. Let $f(x)=(1-e^{-x})/x$. Since $f'(x)=[(1+x)e^{-x}-1]/x^2<0$, it follows that $r_U<0$ and $r_Z>0$. The two limits follow from l'H\^opital's rule and $e^{-x}\to0$.

\subsection{Optimal stopping and the logistic hazard}
Equation~\eqref{eq:Vprivate} is the Bellman equation for a worker who receives exit opportunities at Poisson rate $\chi$ and accepts exactly when $d\geq\widehat V^p$. Therefore $\mu=\chi[1-G(\widehat V^p)]$. For a logistic $G$, direct substitution gives equation~\eqref{eq:mulogit}. Integrating the surplus above the cutoff gives equation~\eqref{eq:optionvalue}. The stopping interpretation does not require the components of $d$ to be separately identified.

\subsection{Proof of Proposition~\ref{prop:socialprotection} and Corollary~\ref{cor:timing}}
Parts (i) and (ii) follow from the stopping rule and equation~\eqref{eq:crowdout}. For part (iii), perturb the displacement flow at date $t$ while holding the destination wage/value path fixed. The marginal worker contributes her own expected transition loss, $L^{\rm own,W}(t)$. Under finite throughput, the additional stock lowers the completion hazard of pre-existing attached workers because $r_U<0$. Holding their exit hazards fixed, a lower completion hazard increases survival in adjustment for affected cohorts and hence their total adjustment-time loss. This component is strictly positive whenever the hazard is reduced over a positive-length interval with positive earnings. The perturbation can also change persistent exit because $\mu$ is decreasing in $\widehat V^p$. That induced-exit component has no general sign in welfare: an additional exit sacrifices $(1-\xi)J^N$ but also reduces subsequent congestion. Collecting the waiting and induced-exit effects gives $\mathcal E^W(t)$ and the identity $\Lambda^W_\Phi=L^{\rm own,W}+\mathcal E^W$. Under proportional throughput $M(U)=\rho_0U$, $r_U=0$, so the perturbation does not change the completion hazard of other workers and the congestion externality is zero. The positive values in Table~\ref{tab:external} are therefore numerical results.

For the corollary, impose the stated equality of private and social technological gains, compare marginal displacement at two dates $t<T$ and let $\Delta\Pi_{tT}$ be the present-value private gain from adopting at $t$ rather than $T$, excluding worker-side transition losses. If the adopter internalizes the marginal worker's own loss, its private threshold for early adoption is $\Delta L_{tT}$, whereas the planner's threshold is $\Delta L_{tT}+\Delta\mathcal E_{tT}=\Delta\Lambda_{tT}$. Hence $\Delta\mathcal E_{tT}>0$ creates the nonempty interval $\Delta L_{tT}\leq\Delta\Pi_{tT}<\Delta\Lambda_{tT}$ in which the adopter chooses $t$ and the planner chooses $T$. If the adopter internalizes none of the worker-side transition loss, its private threshold excludes $\Delta L_{tT}$ and the relevant worker-side wedge is $\Delta\Lambda_{tT}$; a positive value again creates a range of private gains for which early adoption is privately chosen but socially rejected. Because $\mathcal E^W$ and $\Lambda^W_\Phi$ are already date-zero present values, no additional discount factor is required when taking differences across dates. The corollary compares two finite dates and makes no claim about a global optimal diffusion path.

\subsection{Analytical boundaries for adoption speed}
\label{app:frontier}
\subsubsection{Instantaneous adoption}
\label{sec:instantaneous}

Instantaneous adoption provides an analytical upper boundary without choosing an arbitrary maximum value of $\kappa$. As $\kappa\to\infty$, the logistic path jumps from $a_0$ to one at $t=0$. Equivalently, the displacement flow converges weakly to a point mass of size $\bar\imath(1-a_0)$ at zero, so the adjustment system starts from
\begin{equation}
 U(0^+)=\bar\imath(1-a_0),\qquad \Phi(t)=0\quad\text{for }t>0.
 \label{eq:instantstate}
\end{equation}
Because the normalized technology-gain path equals one for every $t>0$ in the limit,
\begin{equation}
 B(\infty)=\frac{1}{\alpha_L}\int_0^\infty e^{-\beta t}dt=\frac{1}{\alpha_L\beta}.
 \label{eq:Binf}
\end{equation}
Let $C^Y(\kappa;Z)=C_D(\kappa;Z)+C_U(\kappa;Z)$ and $C^W(\kappa;Z;\xi)=(1-\xi)C_D(\kappa;Z)+C_U(\kappa;Z)$. For either $q=Y$ or $q=W$,
\[
 \mathcal S_q(\kappa)-\mathcal S_q(\infty)
 =-g_A[B(\infty)-B(\kappa)]+[C^q(\infty;Z)-C^q(\kappa;Z)],
\]
Here $\mathcal S_Y=\mathcal Y$ and $\mathcal S_W=\mathcal W$. Since $B(\infty)>B(\kappa)$, define the preparedness--technology frontier by
\begin{equation}
 g_q^F(Z)=\sup_{\kappa<\infty}
 \frac{C^q(\infty;Z)-C^q(\kappa;Z)}{B(\infty)-B(\kappa)}.
 \label{eq:gF}
\end{equation}
If $g_A<g_q^F(Z)$, at least one finite speed dominates instantaneous adoption under criterion $q$; if $g_A>g_q^F(Z)$, none does. This frontier is only an analytical boundary. The main results compare finite speeds and do not rely on instantaneous adoption.

\subsubsection{Slow adoption and an interior finite speed}
\label{sec:interior}

As $\kappa\to0$, $a(t;\kappa)-a_0\simeq a_0(1-a_0)\kappa t$ and $\Phi(t;\kappa)\simeq\bar\imath a_0(1-a_0)\kappa$ over any fixed discounted horizon. The system is asymptotically uncongested, so $r\to\rho_0$, $J^N\to\bar H$, and $\mu\to\mu_0$. Solving the resulting linear stock equation gives
\[
 B(\kappa)\sim\frac{a_0}{\alpha_L\beta^2}\kappa,\qquad
 C_D(\kappa;Z)\sim\frac{\bar H\mu_0\bar\imath a_0(1-a_0)}{\beta(\beta+\rho_0+\mu_0)}\kappa.
\]
Because finite throughput converges to proportional throughput as the queue vanishes, $C_Q=o(\kappa)$, but ordinary adjustment time enters the first-order cost. In the limit, $w^N\to1$ and
\[
 C_U(\kappa;Z)\sim\frac{\bar w_R\bar\imath a_0(1-a_0)}{\beta(\beta+\rho_0+\mu_0)}\kappa.
\]
Thus the slow-adoption break-even gains under the full-cost criteria are
\begin{align}
 g_Y^L&=\alpha_L\beta\bar\imath(1-a_0)\frac{\bar H\mu_0+\bar w_R}{\beta+\rho_0+\mu_0},\label{eq:gL}\\
 g_W^L&=\alpha_L\beta\bar\imath(1-a_0)\frac{(1-\xi)\bar H\mu_0+\bar w_R}{\beta+\rho_0+\mu_0}.
\end{align}
With $Z$ fixed and continuous objective functions, a gain strictly above the slow boundary and strictly below the fast boundary implies at least one finite maximizing speed: some finite speeds dominate indefinite postponement and some dominate instantaneous adoption, so the maximum occurs away from both limits. Over such long horizons, however, capacity would itself adjust, so this limiting result is not used for policy.

\section{Calibration details}

\subsection{CPS productive-value construction}

The constant $\bar H$ sets the scale for the normalized continuation value, $\widehat V^p=V^p/\bar H$. Actual exit losses in equation~\eqref{eq:CD} are not simply the exit mass times $\bar H$; they use the time-varying reallocation value $J^N(t)$ generated by each transition path. Let $s_j$ denote the displaced-worker shares for young, prime-age, and older workers. The calibration uses
\[
(s_y,s_m,s_o)=(0.3777,0.4138,0.2085).
\]
Lost-job weekly earnings by age group are 1,061.84, 1,770.68 and 1,877.91 dollars. The earnings-retention factors are exponentiated survey-weighted mean log ratios of current to lost-job weekly earnings among observations with positive earnings at both dates and a valid loss year; lost-job earnings in these ratios are converted to January 2024 prices using CPI-U. Work factors are the survey-weighted means of $(1-e^{-0.04(67-a_i)})/0.04$, where $a_i$ is age in years. The implementation multiplies the reported lost-job earnings by retention factors $(0.9498,\allowbreak 0.9175,\allowbreak 0.7401)$ and divides by the weekly earnings numeraire of 1,420.07 dollars.

The resulting post-displacement earnings scales, relative to that numeraire, are $(0.7102,\allowbreak 1.1440,\allowbreak 0.9787)$. The discounted remaining-work factors are $(19.7497,\allowbreak 14.6844,\allowbreak 5.8586)$. 

Their share-weighted product gives $\bar H=13.44$ average-worker annual-wage-years.

The exit probabilities $(0.03948,\allowbreak 0.06408,\allowbreak 0.18336)$ yield the aggregate target $0.07966$. The discounted work factors determine the standardized prospective horizons in equation~\eqref{eq:Jimplementation}, not individual retirement dates. The age profile is used to value the post-displacement stream; age is not a separate dynamic state in the model.

\subsection{WIOA timing}
The public-use WIOA extraction retains dislocated workers aged 20--67 who received training. The archived extract contains 90,339 valid spells. All three reported spells are considered; valid duration is 1--3,652 days, with unresolved spells censored at program exit or the reporting-quarter endpoint. The completion rate is the count of recorded completions divided by total observed person-years (days divided by 365.25). The proxy multiplies this rate by the share employed in the second quarter among completed spells with observed binary employment outcomes. Across all retained spells, annual completion-followed-by-Q2-employment proxy hazards are $(1.371473,1.266131,1.068394)$. Their CPS-share-weighted mean is $1.265$ and is used for $\rho_0$. These program hazards anchor uncongested timing; aggregate capacity $Z$ is a separate scenario parameter.

\subsection{UI mapping and stopping-rule calibration}
The UI mapping uses the monthly total unemployment-exit rate $0.171$ reported by \citet{valletta2014}, split into $0.078$ to employment and $0.093$ to nonparticipation. The total rate discounts the additional UI stream, while the labor-force-exit response identifies the attachment margin. \citet[Table~5, Panel~B]{valletta2014} reports a $0.00514$ response to ten additional UI weeks and \citet{valletta2015erratum} establishes its negative sign. With a 52 percent replacement rate, $\bar H=13.44$, and persistence bridge $p_D=0.25$, equations~\eqref{eq:UIvalue}--\eqref{eq:UIslope} give $s_V=2.82$.

Conditional on $\chi$, $(\gamma,\lambda)$ are calibrated to this local slope and the 7.97 percent CPS exit proxy. The reference normalization $\chi=0.293$ gives $(\gamma,\lambda)=(43.78,59.11)$ in normalized-value units. These parameters describe the total private exit value $d$; its transfer and nonmarket components are not separately identified. Repeating the calibration at $\chi=0.586$ and $1.171$ leaves the baseline welfare thresholds nearly unchanged.

The two moments identify the reduced-form exit rule, not its individual components. The CPS supplies the benchmark exit target, and the UI experiment supplies the slope target for the stock-weighted response to a change in the value of remaining attached. Given $\chi$, those moments locate and scale the distribution of $d$. They cannot separately identify cash benefits, leisure, home production, eligibility, application delays, or other ingredients of the private outside option, nor does the stopping decision require them to. Welfare needs one additional input: $\xi$, the real value of nonwork. Because the exit data do not identify it, $\xi$ comes from external evidence.

\subsection{Nonwork valuation}
\citet{maspallais2019} distinguish $\widetilde z$, the value of nonwork relative to predicted pretax market earnings, from $z$, the corresponding object normalized by after-tax marginal product. Table~1 of the published article reports $\widetilde z=0.58$ (standard error $0.04$) for a full-time job; the nationally representative UAS exercise gives $0.61$. Equation~\eqref{eq:J4} is constructed in market-earnings units and the competitive destination sector equates wages with marginal product, so the baseline sets $\xi=0.58$ without an additional tax or wage--marginal-product conversion. The $\xi=0.30$ sensitivity allows for external-validity concerns because the experimental sample consists of unemployed job applicants rather than persistently displaced workers. Mas and Pallais also report a broader preferred opportunity-cost measure of $0.72$ after adding average commuting, forgone-benefit, and child-care costs. Because that object is expressed in their after-tax marginal-product normalization, the paper uses $0.72$ only as an upper benchmark.

\section{Numerical solution and validation}
\label{app:numerics}

For each pair $(\kappa,Z)$, the solver iterates jointly over the transition equations, the destination-wage path that enters $J^N(t)$, and the private stopping value in equation~\eqref{eq:Vprivate}. The time step is 0.025 years. The arrival mass in each step is $\bar\imath[a(t+\Delta t)-a(t)]$, spread uniformly within that step; completion and exit hazards are held at their start-of-step values. The stock update integrates the resulting linear competing-hazard equation exactly within the step. The value $J^N$ uses equation~\eqref{eq:Jimplementation} with trapezoid integration and interpolation at the group-specific horizons. The Bellman equation is solved backwards using a terminal value of zero at year 145. Costs and calibration moments are integrated through year 100; the benefit integral in equation~\eqref{eq:B} is evaluated over the infinite horizon for every speed. A damping factor of 0.10 is used, and the iteration stops when the maximum difference between successive undamped and current continuation values through year 100 is below $10^{-9}$ ($10^{-12}$ for marginal-loss calculations). The replication files report the residuals, mass-balance checks and a smaller-step, longer-horizon comparison.

Marginal costs are calculated by perturbing the displacement flow over a single time step around the peak or the later evaluation date, holding $w^N$ and $J^N$ fixed, and re-solving the stopping and stock equations. The own loss integrates the corresponding cohort's survival and exit flows on the unperturbed paths. The reported derivative uses a pulse of $10^{-6}$ worker-equivalents; the numerical package also reports pulses of $10^{-5}$ and $10^{-4}$ and the proportional-throughput control.

Appendix~\ref{app:frontier} characterizes instantaneous adoption analytically. The numerical analysis is restricted to the finite speeds in the main text; no numerical estimate of the continuous-speed frontier is reported. Halving the step to 0.0125 years and extending the cost horizon to 150 years, with recalibration, changes the baseline pairwise output thresholds by at most 0.048 percentage points, the welfare thresholds by at most 0.032 percentage points, and the simulated exit shares by at most 0.004 percentage points across the ten baseline comparisons.

\subsection{Calibration-load matrix}

Table~\ref{tab:robust} summarizes three calibrations, each evaluated at three common capacity levels; Table~\ref{tab:loadmatrix} shows the full $0.30\to1$ matrix. Each row calibrates the exit rule at a different reference load while matching the same 7.97 percent exit measure and $s_V=2.82$; each column evaluates that rule at a common capacity.

\begin{table}[H]
\centering
\caption{Full calibration-load matrix for $0.30\to1$}\label{tab:loadmatrix}
\small
\resizebox{\textwidth}{!}{%
\begin{tabular}{ccccccc}
\toprule
 & \multicolumn{2}{c}{Evaluation load 0.5} & \multicolumn{2}{c}{Evaluation load 1} & \multicolumn{2}{c}{Evaluation load 2}\\
\cmidrule(lr){2-3}\cmidrule(lr){4-5}\cmidrule(lr){6-7}
Calibration load & $g^Y$ & $g^W(.58)$ & $g^Y$ & $g^W(.58)$ & $g^Y$ & $g^W(.58)$\\
\midrule
0.5 & 0.059 & 0.035 & 0.121 & 0.071 & 0.158 & 0.090\\
1.0 & 0.055 & 0.034 & 0.120 & 0.072 & 0.155 & 0.090\\
2.0 & 0.028 & 0.028 & 0.084 & 0.079 & 0.146 & 0.090\\
\bottomrule
\end{tabular}%
}
\begin{minipage}{.96\textwidth}\footnotesize
Notes: Each row rematches the same benchmark exit share and UI-implied local slope at the stated calibration load, then evaluates that rule at the three absolute capacities corresponding to evaluation loads $0.5$, $1$, and $2$. Welfare uses $\xi=0.58$.
\end{minipage}
\end{table}

\subsection{Nonmarket-value sensitivity}

\begin{table}[H]
\centering
\caption{Welfare thresholds under alternative nonmarket values}\label{tab:xi}
\small
\begin{tabular}{ccccc}
\toprule
& \multicolumn{2}{c}{$0.30\to1$} & \multicolumn{2}{c}{$1\to2$}\\
\cmidrule(lr){2-3}\cmidrule(lr){4-5}
Reference load & $g^W(.30)$ & $g^W(.72)$ & $g^W(.30)$ & $g^W(.72)$\\
\midrule
0.5 & 0.044 & 0.029 & 0.113 & 0.077\\
1.0 & 0.095 & 0.061 & 0.141 & 0.095\\
2.0 & 0.121 & 0.074 & 0.130 & 0.084\\
4.0 & 0.112 & 0.066 & 0.114 & 0.071\\
6.0 & 0.105 & 0.062 & 0.106 & 0.064\\
\bottomrule
\end{tabular}
\begin{minipage}{.92\textwidth}\footnotesize
Notes: The behavioral calibration and exit paths are identical to the baseline $\xi=0.58$. The lower sensitivity $\xi=0.30$ allows for limited external validity of the experimental nonwork valuation. The upper value $\xi=0.72$ uses the broader Mas--Pallais opportunity-cost estimate including average fixed employment and child-care costs; because that estimate uses a different marginal-product normalization, it is treated only as an upper external benchmark.
\end{minipage}
\end{table}

\subsection{Opportunity-rate sensitivity}

\begin{table}[H]
\centering
\caption{Stopping-opportunity normalization at reference capacity}\label{tab:chi}
\small
\begin{tabular}{ccccc}
\toprule
$\chi$ & Interpretation & $g^Y(.30,1)$ & $g^W(.30,1)$ & $g^W(1,2)$\\
\midrule
0.293 & reference & 0.1195 & 0.07237 & 0.11020\\
0.586 & $2\times$ reference & 0.1201 & 0.07220 & 0.10928\\
1.171 & $4\times$ reference & 0.1203 & 0.07213 & 0.10891\\
\bottomrule
\end{tabular}
\begin{minipage}{.92\textwidth}\footnotesize
Notes: $(\gamma,\lambda)$ are recalibrated at each $\chi$ to reproduce the same 7.97 percent exit proxy and local slope $s_V=2.82$.
\end{minipage}
\end{table}

\subsection{Persistence-bridge sensitivity}

\begin{table}[H]
\centering
\caption{Persistence bridge at reference capacity}\label{tab:pd}
\small
\begin{tabular}{ccccc}
\toprule
$p_D$ & UI-implied $s_V$ & $\chi$ & $g^Y(.30,1)$ & $g^W(.30,1)$\\
\midrule
0.15 & 1.69 & 0.176 & 0.116 & 0.074\\
0.25 & 2.82 & 0.293 & 0.120 & 0.072\\
0.35 & 3.95 & 0.410 & 0.122 & 0.071\\
\bottomrule
\end{tabular}
\begin{minipage}{.92\textwidth}\footnotesize
Notes: For each $p_D$, the opportunity-rate normalization is $\chi=p_Dq_0$ and $(\gamma,\lambda)$ are recalibrated to match the 7.97 percent CPS exit proxy and the corresponding UI-implied slope in equation~\eqref{eq:UIslope}. Welfare uses $\xi=0.58$. The behavioral anchor absorbs most of the variation in the persistence bridge.
\end{minipage}
\end{table}

\subsection{Destination absorption and behavioral recalibration}

\begin{table}[H]
\centering
\caption{Destination absorption at reference capacity}\label{tab:sigma}
\small
\begin{tabular}{ccccc}
\toprule
$\sigma$ & $g^Y(.30,1)$ & $g^W(.30,1)$ & $g^Y(1,2)$ & $g^W(1,2)$\\
\midrule
1.5 & 0.120 & 0.072 & 0.175 & 0.110\\
3.0 & 0.136 & 0.077 & 0.198 & 0.115\\
6.0 & 0.145 & 0.080 & 0.213 & 0.119\\
\bottomrule
\end{tabular}
\begin{minipage}{.94\textwidth}\footnotesize
Notes: All entries use $Z=2.63$ percent per year. For each $\sigma$, the stopping rule is recalibrated to the same 7.97 percent exit share and $s_V=2.82$ at $\kappa=0.30$. Welfare uses $\xi=0.58$.
\end{minipage}
\end{table}

\subsection{Additional robustness tables}

The main-text robustness table condenses the following exercises for capacity growth, calibration load, throughput, and the automation envelope.

\begin{table}[H]
\centering
\caption{Exogenous growth of adjustment capacity at reference initial capacity}\label{tab:zgrowth}
\small
\resizebox{\textwidth}{!}{%
\begin{tabular}{ccccccc}
\toprule
$\eta$ & \multicolumn{2}{c}{$g^Y(0.30,1)$} & \multicolumn{2}{c}{$g^W(0.30,1)$} & \multicolumn{2}{c}{$g^W(1,2)$}\\
\cmidrule(lr){2-3}\cmidrule(lr){4-5}\cmidrule(lr){6-7}
 & fixed & recal. & fixed & recal. & fixed & recal.\\
\midrule
0\% & 0.120 & 0.120 & 0.072 & 0.072 & 0.110 & 0.110\\
2\% & 0.115 & 0.115 & 0.070 & 0.070 & 0.115 & 0.116\\
5\% & 0.105 & 0.107 & 0.065 & 0.065 & 0.122 & 0.122\\
\bottomrule
\end{tabular}%
}
\begin{minipage}{.96\textwidth}\footnotesize
Notes: $Z_0=2.625$ percent of the workforce per year and $Z(t)=Z_0e^{\eta t}$. ``Fixed'' holds the reference stopping rule fixed; ``recal.'' rematches the 7.97 percent exit share and $s_V=2.82$ at $\kappa=0.30$ for each capacity-growth rate. Welfare uses $\xi=0.58$.
\end{minipage}
\end{table}

\begin{table}[H]
\centering
\caption{Calibration-load and throughput robustness}\label{tab:robust}
\small
\resizebox{\textwidth}{!}{%
\begin{tabular}{ccccccc}
\toprule
$Z$ (\%/yr) & Ref. load & $g^Y$ ref. & $g^Y$ load band & $g^W$ ref. & $g^W$ load band & MM $g^W$\\
\midrule
5.25 & 0.5 & 0.055 & 0.028--0.059 & 0.034 & 0.028--0.035 & 0.047\\
2.63 & 1.0 & 0.120 & 0.084--0.121 & 0.072 & 0.071--0.079 & 0.074\\
1.31 & 2.0 & 0.155 & 0.146--0.158 & 0.090 & 0.090 & 0.079\\
\bottomrule
\end{tabular}%
}
\begin{minipage}{.96\textwidth}\footnotesize
Notes: The load band recalibrates the same 7.97 percent exit share and local slope at reference loads $0.5$, $1$, and $2$, then evaluates the calibrations at the same absolute $Z$. The Michaelis--Menten case uses $M(U,Z)=\rho_0U/[1+\rho_0U/Z]$ and rematches the same behavioral moments. Welfare uses $\xi=0.58$.
\end{minipage}
\end{table}

\begin{table}[H]
\centering
\caption{Threshold sensitivity to the automation envelope}\label{tab:ibar}
\small
\resizebox{\textwidth}{!}{%
\begin{tabular}{ccccccc}
\toprule
 & \multicolumn{3}{c}{Market output $g^Y(0.30,1)$} & \multicolumn{3}{c}{Welfare $g^W(0.30,1)$}\\
\cmidrule(lr){2-4}\cmidrule(lr){5-7}
$\bar\imath$ & $Z=5.25\%$ & $Z=2.63\%$ & $Z=1.31\%$ & $Z=5.25\%$ & $Z=2.63\%$ & $Z=1.31\%$\\
\midrule
0.25 & 0.015 & 0.048 & 0.108 & 0.013 & 0.038 & 0.067\\
0.35 & 0.055 & 0.120 & 0.155 & 0.034 & 0.072 & 0.090\\
0.45 & 0.100 & 0.174 & 0.196 & 0.055 & 0.099 & 0.111\\
\bottomrule
\end{tabular}%
}
\begin{minipage}{.94\textwidth}\footnotesize
Notes: The behavioral rule and absolute $Z$ values are held fixed; only the terminal automation envelope varies. The external gain scenarios can be scaled in proportion to $\bar\imath/0.35$ when comparing gain per automated unit.
\end{minipage}
\end{table}

\subsection{Ordinary-training timing}
The full adjustment-time loss enters both output and welfare. At reference capacity, the uncongested timing component adds 0.0041 to the $0.30\to1$ threshold and 0.0040 to the $1\to2$ threshold. The rest of the adjustment-time component is the net difference relative to the counterfactual with proportional throughput, re-solving wages and exit behavior.

\section{Secondary wage implication}
\label{app:wage}

Because the destination sector has its own technology, its wage follows the inverse labor-demand schedule in equation~\eqref{eq:w4}. A faster inflow of successfully retrained workers initially lowers $w^N$. If the faster path later produces enough exit to leave fewer workers in the destination sector, the wage ordering reverses. This crossing is a feature of the transition; the thresholds are computed using the full wage paths.

\section{Notation}
\begin{table}[H]
\centering
\caption{Main notation}
\small
\begin{tabular}{>{\raggedright\arraybackslash}p{.18\textwidth} p{.70\textwidth}}
\toprule
Symbol & Meaning\\
\midrule
$a(t;\kappa)$ & Fraction of the fixed automation envelope adopted at date $t$\\
$\kappa$ & Adoption speed\\
$a_0$ & Initial adoption share\\
$\bar\imath$ & Terminal share of labor tasks displaced\\
$\Phi(t)$ & Flow of newly displaced workers\\
$U(t)$ & Stock of displaced workers attached to adjustment\\
$Z$ & Installed aggregate adjustment capacity\\
$\rho_0$ & Uncongested completion hazard\\
$M(U;Z)$ & Aggregate successful-adjustment throughput\\
$r(U,Z)=M/U$ & Individual completion hazard\\
$w^N(t)$ & Destination-sector wage index ($w_0^N=1$)\\
$J^N(t)$ & Discounted productive value after successful reallocation\\
$\widehat V^p$ & Private continuation value normalized by $\bar H$\\
$\mu(t)$ & Persistent-exit hazard\\
$\chi$ & Arrival rate of exit opportunities\\
$d$ & Private value of supported nonparticipation\\
$\xi$ & Social value of nonwork relative to the forgone productive stream\\
$C_D$ & Market-output loss from persistent exit\\
$C_U$ & Total adjustment-time loss\\
$C_U^0$ & Adjustment-time loss under proportional throughput\\
$C_Q$ & Net excess adjustment-time loss relative to uncongested adjustment\\
$B(\kappa)$ & Discounted benefit from reaching the technology frontier earlier\\
$g_A$ & Remaining level gain in labor productivity from initial to full adoption\\
$g^W,\ g^Y$ & Pairwise technological-gain thresholds for welfare and market output\\
$\Lambda^W_\Phi$ & Marginal welfare cost of displacement\\
$L^{\rm own,W}$ & Marginal displaced worker's own transition loss\\
$\mathcal E^W$ & Congestion externality on other users of adjustment capacity\\
\bottomrule
\end{tabular}
\end{table}

\end{document}